\documentstyle[11pt]{article}

\def\dalemb#1#2{{\vbox{\hrule height .#2pt
        \hbox{\vrule width.#2pt height#1pt \kern#1pt
                \vrule width.#2pt}
        \hrule height.#2pt}}}
\def\square{\mathord{\dalemb{5.8}{6}\hbox{\hskip1pt}}}

\let\a=\alpha \let\b=\beta   \let\e=\epsilon

 \def\bd{\begin{document}} \def\ed{\end{document}}
\def\ds{\documentstyle} \let\fr=\frac \let\bl=\bigl \let\br=\bigr
\let\Br=\Bigr \let\Bl=\Bigl 
\let\bm=\bibitem
\let\na=\nabla
\let\pa=\partial \let\ov=\overline
\let\ul=\underline 
\newcommand{\be}{\begin{equation}} 
\newcommand{\ee}{\end{equation}} 
\def\ve{\varepsilon}
\def\ba{\begin{array}}
\def\ea{\end{array}}
\def\ft#1#2{{\textstyle{{\scriptstyle #1}\over {\scriptstyle #2}}}}
\def\fft#1#2{{#1 \over #2}}
\def\del{\partial}
\def\sst#1{{\scriptscriptstyle #1}}
\def\oneone{\rlap 1\mkern4mu{\rm l}}
\def\e7{E_{7(+7)}}
\def\td{\tilde}
\def\bog{Bogomol'nyi\ }
\def\ads{anti-de Sitter\ }
\newcommand{\ho}[1]{$\, ^{#1}$}
\newcommand{\hoch}[1]{$\, ^{#1}$}
\newcommand{\bea}{\begin{eqnarray}} 
\newcommand{\eea}{\end{eqnarray}} 
\newcommand{\ra}{\rightarrow}
\newcommand{\lra}{\longrightarrow}
\newcommand{\Lra}{\Leftrightarrow}
\newcommand{\ap}{\alpha^\prime}
\newcommand{\bp}{\tilde \beta^\prime}
\newcommand{\tr}{{\rm tr} }
\newcommand{\Tr}{{\rm Tr} } 
\newcommand{\NP}{Nucl. Phys. }
\newcommand{\tamphys}{\it $^{(1)}$ Center for Theoretical Physics,
Texas A\&M University, College Station, Texas 77843}

\newcommand{\auth}{H. L\"u$^{(1,2)}$, C.N. Pope$^{(1,2)}$ and 
P.K. Townsend$^{(3)}$ }

\begin{document}
\begin{flushright}
\hfill{CTP TAMU-26/96}\\
\hfill{SISSA 109/96/EP}\\
\hfill{hep-th/9607164}\\
\end{flushright}

\vspace{20pt}

\begin{center}
{\large {\bf Domain Walls from Anti-de Sitter Spacetime}}

\vspace{30pt}

\auth

\vspace{15pt}

{\tamphys}

\vspace{10pt}

{\it $^{(2)}$ SISSA, Via Beirut No. 2-4, 34013 Trieste, Italy }

\vspace{10pt}

{\it $^{(3)}$ DAMTP, University of Cambridge, 
Silver St., Cambridge CB3 9EW, U.K.}



\vspace{40pt}

\underline{ABSTRACT}
\end{center}

     We examine $(D-2)$-brane solutions in supergravities, showing that
they fall into four categories depending on the details of the dilaton
coupling.  In general they describe domain walls, although in one of the
four categories the metric describes anti-de Sitter spacetime.  We study
this case, and its $S^1$ dimensional reduction to a more conventional 
domain wall in detail, focussing in particular on the manner in which the 
unbroken supersymmetry of the anti-de Sitter solution is partially broken 
by the dimensional reduction to the domain wall.

{\vfill\leftline{}\vfill
\vskip	10pt
\footnoterule
{\footnotesize
	Research supported in part by DOE 
Grant DE-FG05-91-ER40633 and \vskip	-12pt}  \vskip	10pt
{\footnotesize 
      EC Human Capital and Mobility Programme under contract ERBCHBGCT920176.  
} 
}

\pagebreak
\setcounter{page}{1}     

\section{Introduction}

     The past few years have witnessed an exhaustive study of $p$-brane
solutions of $D$-dimensional supergravity theories, motivated by their
increasing importance in the non-perturbative dynamics of superstring
theories, or M-theory, compactified to $D$ dimensions. A special class of
these solutions are the $(D-2)$-branes, which can be viewed as domain walls.
$(D-2)$-branes exhibit some features that are significantly different from
those of more generic $p$-brane solutions and their properties are
consequently much less well known, with the exception of domain walls in
$D=4$, $N=1$ supergravity which have been the subject of extensive
investigations (see \cite{cs} for a recent  review of this work and further
references). Rather than the usual situation where a propagating
antisymmetric tensor field carries a magnetic or electric charge that
supports the $p$-brane, $(D-2)$-branes are supported either by a
cosmological-type term in the Lagrangian, or else, in the dualised version,
by a non-propagating $D$-form field strength. The necessary terms arise in
massive supergravity theories and in gauged supergravities.  A case that has
recently attracted  attention is the massive type IIA supergravity in $D=10$
\cite{r}, which  admits an 8-brane solution \cite{pw}.  Other examples,
involving massive gauged supergravities in $D=7$ and $D=6$, have been
discussed in \cite{lpss2}.  In all these cases, the domain wall solutions
preserve half the supersymmetry.  In this paper, we shall discuss the
general structure of $(D-2)$-branes preserving half the supersymmetry of the
maximal supergravity in general dimension $D$. As we shall see, they fall
naturally into four categories according to the type of
supergravity theory. 

    One category, let us call it the first, arises from gauged supergravity
theories. We shall have little to say about this case here. The second
category arises from so called `massive' supergravity theories. Numerous
examples are provided by `generalized' dimensional reduction of $D=11$
supergravity but an earlier example of current interest, which cannot be
obtained in this way, is the massive IIA supergravity in $D=10$, which
admits an 8-brane solution. It was shown in \cite{bdgpt} that the massive
type IIA supergravity can be dimensionally reduced to a massive $N=2$
supergravity in $D=9$; the 8-brane solution in $D=10$ can then be reduced to
a 7-brane solution of this $D=9$ theory. At first sight the reduction of the
{\it massive} $D=10$ theory to $D=9$ might seem surprising since it is often
said that Kaluza-Klein dimensional reduction is possible only if the
higher-dimensional theory admits an $S^1\times M_{\sst D-1}$ direct-product
vacuum solution, which the massive IIA theory does not.  However, it was
argued in \cite{bdgpt} that it was sufficient that the 8-brane solution has
a translational invariance, which, after compactification, corresponds to a
$U(1)$ invariance of the solution.  In fact, as we shall discuss in section 
3, neither this, nor indeed any other solution, plays any role in the 
dimensional reduction of the theory.

    The observation that an $S^1\times M_{\sst D-1}$ direct-product vacuum
is unnecessary for Kaluza-Klein reduction plays a crucial role in
understanding the third category of $(D-2)$-branes, which provide the main
focus of this paper. These can be understood as arising from dimensional
reduction of anti-de Sitter (AdS) space in $(D+1)$ dimensions.  Anti-de
Sitter space itself can be viewed, in horospherical coordinates, as a
special case of a $(D-1)$-brane in $(D+1)$ dimension, {\it i.e.}\ a domain
wall, in which the manifest $D$-dimensional Poincar{\' e} isometry is
`accidentally' enlarged.  This can be double-dimensionally reduced, just
like any other $p$-brane, to give a domain wall solution in $D$ dimensions.
There is a significant difference, however. Just as AdS space has an
`accidentally' enlarged bosonic symmetry it also has, in the supergravity
context, an `accidentally' enlarged supersymmetry since, unlike typical
$p$-brane solutions which break half the supersymmetries, the AdS solution
breaks none. This is puzzling in view of the fact that the domain wall
solution in the lower dimension certainly breaks half the supersymmetry of
the dimensionally reduced supergravity theory. The resolution of this puzzle
is that only half the Killing spinors in the AdS spacetime are independent
of the compactified coordinate, and hence the other half are lost in the
dimensional reduction. We demonstrate this by giving an explicit
construction of all the Killing spinors in the $D$-dimensional AdS
spacetime, in the natural (horospherical) coordinate system that arises in
its construction as a $(D-2)$-brane. 

     In dimensionally reducing a $p$-brane solution in $(D+1)$ dimensions to
one in $D$ dimensions one must first periodically identify one of the
$p$-brane coordinates. From what we have just said, it is clear that this
can be done for $(D+1)$-dimensional AdS space too. If this is done, the
periodic coordinate can be re-interpreted as the azimuthal angle of a
circularly-symmetric domain wall solution in $(D+1)$-dimensions. In
particular, for $D=2$, {\it i.e.}\ $(D+1)=3$, we have a particle-like
solution of the Einstein equations with a cosmological constant that is {\it
locally} isometric to ${\rm AdS}_3$ (as are all solutions in $2+1$
dimensions) and which preserves at least half the supersymmetry. In fact,
the solution preserves just half the supersymmetry because the other half is
broken by the identification; this solution is just the supersymmetric
`black hole vacuum' discussed in \cite{hc} and exhibited in horospherical
coordinates in \cite{it}. Thus, the periodic identification
of \ads space as described above leads to a straightforward generalization
to arbitrary dimension of the $2+1$ dimensional `black hole vacuum'. An
interesting question, not adressed here, is whether there is also an
analogue in arbitrary dimension of the extreme spinning black hole of $2+1$
gravity \cite{btz}, which is also locally isometric to \ads, and is
supersymmetric. 

     In the following section we shall introduce the four categories of 
$(D-2)$ branes and give some of their properties, concentrating on the
$p$-brane interpretation of \ads space. We then show how $(D-2)$-brane
solutions in D-dimensions are obtained by dimensional reduction of AdS in
$(D+1)$ dimensions and elucidate the supersymmetry of these solutions. 

\section{Solitonic $(D-2)$-branes}

      In general, there exist $p$-brane solutions in supergravity theories
that involve the metric tensor, a dilaton field, and an antisymmetric tensor
field strength of rank $n$.  If the field strength carries an electric
charge, the corresponding elementary $p$-brane has $p=n-2$, whereas if the
charge is magnetic, the $p$-brane is solitonic with $p=D-n-2$, where $D$ is
the spacetime dimension of the theory.  Included in the general solitonic
cases is the degenerate case where $n=0$, which implies that there is no
field strength at all, but instead a cosmological-type term in the
$D$-dimensional effective Lagrangian: 
\be
{\cal L} = e R -\ft12 e (\del\phi)^2 + 2e\Lambda e^{-a\phi} \ .
\label{lag1}
\ee
It is useful to parameterise the constant $a$ in terms of $\Delta$, defined
by
\be
a^2=\Delta +\fft{2(D-1)}{D-2}\ .\label{aval}
\ee
The equations of motion admit solitonic $(D-2)$-brane solutions, 
with metric and dilaton given by \cite{lpss1}
\bea
ds^2 &=& H^{\ft4{\Delta (D-2)}}\, \eta_{\mu\nu}\, dx^\mu\, dx^\nu 
    + H^{\ft{4(D-1)}{\Delta(D-2)}}\, dy^2\ ,\label{dwmetric}\\
e^\phi &=& H^{\ft{2a}\Delta} \, \label{dwdil}
\eea
where $H$ is an harmonic function on the 1-dimensional transverse space, of
the form $H \sim c \pm m y$, where $c$ is an arbitrary integration constant
and $m=\sqrt{-\Lambda\Delta}$.  The arbitrariness of the sign in $H$ arises
because the equations of motion involve $m$ quadratically.  (The solutions 
(\ref{dwdil}) in $D=4$ were also obtained in \cite{c}.)  In section 4, we
shall discuss the elementary $(D-2)$-brane using a $D$-form field strength. 

     A metric of the form $ds^2=H^{2\a} \, \eta_{\mu\nu}\, dx^\mu\, 
dx^\nu + H^{2\b}\, dy^2$, in the vielbein basis $e^{\ul\mu}=H^\a\, dx^\mu$,
$e^{\ul y} = H^\b \, dy$, has a curvature 2-form given by
\bea
\Theta^{\mu\nu}&=&-\a^2\, H^{-2\b-2} H'^2\, e^{\ul \mu}\wedge
e^{\ul \nu}\ ,\nonumber\\
\Theta^{\ul\mu \ul y} &=&\Big( \a(\b-\a+1) H^{-2\b -2}\, H'^2 -
\a H^{-2\b-1}\, H'' \Big) e^{\ul\mu} \wedge e^{\ul y}\ ,\label{dwcurv}
\eea
where $(\ul\mu,\ul y)$ denote tangent-space components.  Thus we can see
that the  solitonic $(D-2)$-brane solutions (\ref{dwmetric}) fall into 
three different 
categories, depending on the value of $\Delta$, according to whether 
\be
\b+1=\fft{2(D-1)}{\Delta (D-2)} + 1 \label{crit}
\ee
is positive, negative, or zero, which determines whether the
curvature diverges at $H=0$, $H=\infty$, or nowhere, 
respectively.\footnote{There is also a fourth category, where $\Delta=0$, 
for which the solutions would be transcendental.  $\Delta=0$ seems never to 
occur in supergravity theories, and we shall not consider this fourth 
category further.}
As we shall discuss in more detail presently, in this latter case, namely
$\Delta=\Delta_{\rm AdS}$, where
\be
\Delta_{\rm AdS}=-\fft{2(D-1)}{D-2}\ ,\label{adsdel}
\ee
corresponding to $a=0$, the metric (\ref{dwmetric}) describes \ads 
spacetime.  If (\ref{crit}) is positive, which 
corresponds to $\Delta >0$, the 
singularity that would occur at $H=0$ can be avoided by taking $H$ to
be
\be
H=1 +m|y|\ .\label{dwh}
\ee
In this case, the metric is free of curvature singularities except for a
delta-function in the curvature at $y=0$, as can be seen from
(\ref{dwcurv}). The metric is asymptotically flat as $y\rightarrow
\pm\infty$, in which regions the dilaton diverges. The solution describes a
domain wall across the spacetime, located at $y=0$. If the constant 1 were
omitted in (\ref{dwh}), the curvature would diverge as some inverse power of
$y$ at $y=0$, where the domain wall is located. The third category of
solution arises when (\ref{crit}) is negative, which corresponds to $
\Delta_{\rm AdS} <\Delta < 0$.  In this case, the singularity at $H=\infty$
cannot be avoided by any choice of the constant of integration.  The choice
(\ref{dwh}) for $H$ ensures that the dilaton remains real, and finite for
finite $y$; the metric then describes a domain wall at $y=0$, embedded in a
spacetime whose curvature and dilaton diverge as $y\rightarrow \pm\infty$.
On the other hand if the constant term in (\ref{dwh}) is omitted, the region
$y\rightarrow 0$ is asymptotically flat, but now the dilaton diverges at
$y=0$ as well as when $y\rightarrow\pm\infty$. 

    All three categories of $(D-2)$-brane solutions can arise in
supergravity theories.  The case $\Delta=\Delta_{\rm AdS}$ occurs, for
example, in certain vacua of gauged supergravities, such as arise from the
$S^7$ \cite{dp} or $S^4$ \cite{pnt} compactifications of $D=11$
supergravity, or the $S^5$ \cite{krn} compactification of $D=10$ type IIB
supergravity. Examples with positive $\Delta$ include the massive type IIA
theory in $D=10$ \cite{r}, and all its dimensional reductions, which all
have $\Delta=4$.  On the other hand, the cosmological-type terms associated
with the gauging of the $D=7$ \cite{tn} and $D=6$ \cite{r2} supergravities
discussed in \cite{lpss2} have $\Delta=-2$, satisfying the inequalities
$\Delta_{\rm AdS} < \Delta <0$ for the third category of $(D-2)$-branes
described above. 

    Let us now consider the special case $\Delta=\Delta_{\rm AdS}$ ({\it
i.e.}\ $a=0$) in more detail.  The metric (\ref{dwmetric}) becomes 
\be
ds^2 = H^{-\ft2{D-1}}\, \eta_{\mu\nu}\, dx^\mu \, dx^\nu + 
H^{-2}\, dy^2\ ,\label{ymetric}
\ee
where as in all the cases discussed above, $H$ has the general form
$H\sim c \pm my$.  The coordinate transformation 
\be
H=\, e^{-mr} \ ,\label{exp}
\ee
puts the metric into the form
\be
ds^2 = e^{2\lambda r}\, \eta_{\mu\nu}\, dx^\mu \, dx^\nu + dr^2\ ,
\label{ads}
\ee
where 
\be
\lambda= \sqrt{\ft{2\Lambda}{(D-1)(D-2)}}= (D-1) m\ .\label{lam}
\ee
This is in fact the metric of
\ads spacetime, in horospherical coordinates \cite{dgt}.  This can be seen
by introducing the $(D+1)$ coordinates $(X,Y,Z^\mu)$
defined by
\bea
X&=& \fft1{\lambda}\, \cosh\lambda r +\ft12\lambda\, \eta_{\mu\nu}\, 
x^\mu x^\nu  \, 
e^{\lambda r} \ ,\nonumber\\
Y&=& \fft1{\lambda}\, \sinh\lambda r -\ft12\lambda\, \eta_{\mu\nu}\, 
x^\mu x^\nu \,
e^{\lambda r} \ ,\label{embed}\\
Z^\mu&=&x^\mu\, e^{\lambda r}\ .\nonumber
\eea
They satisfy 
\bea
\eta_{\mu\nu}\, Z^\mu Z^\nu +Y^2 -X^2 &=&-1/\lambda^2\ ,\label{emcon}\\
\eta_{\mu\nu}\, dZ^\mu dZ^\nu +dY^2 -dX^2&=& 
e^{2\lambda r} \eta_{\mu\nu}\, dx^\mu\, dx^\nu + dr^2\ ,
\eea
which shows that (\ref{ads}) is the induced metric on the hyperboloid 
(\ref{emcon}) embedded in a flat $(D+1)$-dimensional spacetime with
$(-,+,+,\cdots, +,-)$ signature.  Thus (\ref{ads}) has an $SO(2,D-1)$ 
isometry, and it is a metric on $D$-dimensional \ads spacetime.  

     The global structure of the AdS metric is of course well known.  The
structure for the metric (\ref{ads}) in horospherical coordinates was
discussed in the case of $D=4$ dimensions in \cite{dgt}; the situation here
for arbitrary $D$ is similar.  It is evident from (\ref{embed}) that
$X+Y=\lambda^{-1}\, e^{\lambda r}$ is non-negative if $r$ is real, and hence
the region $X+Y<0$ in the full \ads spacetime is not covered by the
horospherical coordinates. In fact, the coordinates used in (\ref{ads})
cover one half of the complete \ads spacetime, and the metric describes
AdS$_D /J$ where $J$ is the antipodal involution $(X,Y, Z^\mu)\rightarrow
(-X, -Y, -Z^\mu)$ \cite{dgt}. If $D$ is even, we can extend the metric
(\ref{ymetric}) to cover all of \ads spacetime by taking $H= my$, since now
the region with $y<0$ corresponds to the previously inaccessible region
$X+Y<0$.  On the other hand if $D$ is odd, we must restrict $H$ in
(\ref{ymetric}) to be non-negative in order to have a real metric, and thus
in this case we should choose $H=c + m |y|$, with $c\ge0$.  If the constant
$c$ is zero, the metric describes AdS$_D/J$, while if $c$ is positive, the
metric describes a smaller portion of the complete \ads spacetime.  In any
dimension, if we have $H=c +m |y|$ (by choice if $D$ is even, or by
necessity if $D$ is odd), the solution can be interpreted as a domain wall
at $y=0$ dividing two portions of \ads spacetimes, with a delta function
curvature singularity at $y=0$ if the constant $c$ is positive.

\section{Domain walls from AdS spacetime}

\subsection{Dimensional reduction with a cosmological term}

     In general, a $p$-brane solution in $D$ dimensions can be dimensionally
reduced to a $(p-1)$-brane solution in $(D-1)$ dimensions, by using the
Kaluza-Klein procedure.  In the case of a $(D-2)$-brane, this might seem
problematical at first sight, since a direct product metric on $S^1\times
M_{\sst D-1}$ is not a solution of the equations of motion following from
(\ref{lag1}) (at least for finite $\phi$).  However, it was shown in
\cite{bdgpt} that the dimensional reduction of massive type IIA supergravity
in $D=10$ \cite{r} to massive $N=2$ supergravity in $D=9$ is nevertheless
possible.  In this case, the 8-brane of the $D=10$ theory \cite{pw} reduces
to a 7-brane of the $D=9$ theory.  The ability to perform the
dimensional reduction was attributed in \cite{bdgpt} to the existence of the
8-brane solution, which has a $U(1)$ isometry.  In fact even the
existence of this solution with a $U(1)$ isometry is inessential for the
Kaluza-Klein reduction of the theory, since dimensional reduction is a
procedure that is applied at the level of the Lagrangian, without making
reference to any particular solution.  A consistent Kaluza-Klein dimensional
reduction can always be performed on any Lagrangian, simply by restricting
the higher-dimensional fields to be independent of the chosen
compactification coordinate.  The solutions of the resulting
lower-dimensional theory will be in one-to-one correspondence with solutions
of the higher-dimensional theory that admit a translational isometry.
Knowing of the existence of a particular such solution of the
higher-dimensional theory provides an assurance that the lower-dimensional
theory is not empty.

     We shall give a more striking illustration of the fact that consistent
Kaluza-Klein reduction is possible in the absence of an $S^1\times M_{\sst
D-1}$ vacuum solution by considering the dimensional reduction of the
Lagrangian (\ref{lag1}) in the case when $a=0$, showing that the $S^1$
reduction of Einstein gravity even with a pure cosmological constant is also
possible.  Let us begin with a Lagrangian of the form (\ref{lag1}), with
$\phi=0$, 
\be
{\cal L}= e R +2e \Lambda \ .\label{lag2}
\ee  
The standard Kaluza-Klein ansatz for the metric is
\be
d s^2= e^{2\alpha\varphi}\, d\tilde s^2 + e^{-2(D-3)\a\varphi}\,
(dz +{\cal A})^2\ ,\label{kkans}
\ee
where $d\tilde s^2$ denotes the $(D-1)$-dimensional metric, and 
$\a^2=(2(D-2)(D-3))^{-1}$.  The tangent-space components of the
$D$-dimensional Ricci tensor $R{\sst A\sst B}$ are
given by
\bea
R_{\tilde {\sst A}\tilde {\sst B}} &=& e^{-2\alpha\varphi} \Big(
\widetilde R_{\tilde{\sst A}\tilde\sst B} -
\ft12 \del_{\tilde\sst{A}}\varphi\, \del_{\tilde\sst{B}}\varphi -
\alpha\,  \widetilde{\square}\varphi \, \eta_{\tilde\sst A\tilde 
\sst B}\Big) -
\ft12 e^{-2(D-1)\alpha\varphi}\, 
{\cal F}_{\tilde \sst{A}}{}^{\tilde \sst{C}} {\cal F}_{\tilde \sst{
B}\tilde\sst{C}}\ ,\nonumber\\
R_{{\tilde \sst{A}}\underline z} &=& \ft12 e^{(D-4)\a\varphi}\,
\widetilde\nabla^{\tilde\sst{B}}\Big( e^{-2(D-2)\a\varphi}\, 
{\cal F}_{\tilde\sst{A}\tilde\sst{B}} \Big)\ ,
\label{ricten}\\
R_{\underline z \underline z} &=& (D-3)\, \alpha\, e^{-2\alpha\varphi}\,
\widetilde{\square} \varphi+\ft14 e^{-2(D-1)\alpha\varphi}\, 
{\cal F}^2\ ,\nonumber
\eea
and thus the equations of motion for the $(D-1)$-dimensional fields,
obtained by substituting into the $D$-dimensional equations of motion
$R_{\sst A \sst B}-\ft12 R \eta_{\sst A \sst B}=
\Lambda \eta_{\sst A \sst B}$ are (after converting to world indices)
\bea
&&\widetilde R_{\tilde\sst{M}\tilde\sst{N}} -
\ft12 \widetilde R \tilde g_{\tilde\sst{M}\tilde\sst{N}}-
\ft12 \del_{\tilde\sst M}\varphi\, 
\del_{\tilde\sst N}\varphi +\ft14 \tilde g_{\tilde\sst{M}\tilde\sst{N}}
(\del\varphi)^2\nonumber\\
&&\qquad-\ft12 e^{-2(D-2)\a\varphi} 
({\cal F}^2_{\tilde\sst{M}\tilde\sst{N}} -\ft14 {\cal F}^2
\tilde g_{\tilde\sst{M}\tilde\sst{N}}) -
\Lambda \, \tilde g_{\tilde\sst{M}\tilde\sst{N}}\, 
e^{2\a\varphi} =0\ ,\nonumber\\
&&\widetilde\nabla_{\tilde\sst M} 
\Big(e^{-2(D-2)\a\varphi}\, {\cal F}^{\tilde\sst{M}\tilde\sst{N}}
\Big) =0\ ,\label{eqs}\\
&&\widetilde{\square}\varphi +\ft12(D-2)\a\, 
e^{-2(D-2)\a\varphi} {\cal F}^2 +
4\a\Lambda\, e^{2\a\varphi} =0\ .\nonumber
\eea
On the other hand, substituting the Kaluza-Klein ansatz into the 
$D$-dimensional Lagrangian (\ref{lag2}) gives the $(D-1)$-dimensional
Lagrangian
\be
{\cal L} = \tilde e \widetilde R -\ft12 \tilde e\, (\del\varphi)^2 
-\ft14 \tilde e\, e^{-2(D-2)\a\varphi}\,
{\cal F}^2 + 2 \tilde e \Lambda \,e^{2\a\varphi}
\ ,\label{lag3}
\ee
whose equations of motion give precisely (\ref{eqs}).  That the terms
independent of $\Lambda$ should agree is of course unremarkable, since these
correspond just to the dimensional reduction of gravity without a
cosmological constant.  It is not always appreciated, however, that the
reduction with the cosmological constant, where there is no $S^1\times M$
vacuum solution, is also consistent. 

    The $(D-2)$-brane solution of the $D$-dimensional theory (\ref{lag2})
is, as we saw earlier, \ads spacetime, which has translational isometries in
the world-volume directions.  Thus we can perform the above Kaluza-Klein
dimensional reduction with the extra coordinate $z$ taken to be one of the
spatial world-volume coordinates $x^i$.  The resulting $(D-1)$-dimensional
metric is nothing but the metric (\ref{dwmetric}), describing a domain wall,
with the same value of $\Delta$ as in $D$ dimensions, corresponding to the
third category described in section 2, where (\ref{crit}) is negative. This
preservation of $\Delta$ under dimensional reduction is also observed for
the usual $p$-brane solutions \cite{lpss1}.  In terms of the dilaton
coupling constant $a$, we have gone from a theory with $a=0$ in $D$
dimensions to one with $a=-2\a$ in $(D-1)$ dimensions.  Note that the
singular curvature of the lower-dimensional domain-wall solution arises from
the dimensional reduction, since the \ads spacetime in the higher dimension
is free of singularities.  A similar phenomenon was discussed in \cite{ght}
for the usual kinds of $p$-branes. 

\subsection{Supersymmetry}

     A curious feature of the above dimensional reduction emerges if we
consider it in the context of supergravity.  Thus let us consider the case
of a $D$-dimensional gauged supergravity theory with a scalar potential
which, for some suitable restriction of the fields, gives a bosonic
Lagrangian of the form (\ref{lag2}).  Upon Kaluza-Klein reduction, we expect
that the theory will yield a supergravity theory in $(D-1)$ dimensions with
the same number of components of unbroken supersymmetry. However, the \ads
solution in $D$ dimensions, which preserves all the supersymmetry, reduces
to the domain-wall solution in $(D-1)$ dimensions, which preserves only half
of the supersymmetry.  To understand where the other half of the
supersymmetry is lost, we need to look at the detailed forms of the Killing
spinors in the \ads spacetime. 

     Let us first calculate the spin connection for the metric (\ref{ads}). 
We begin by choosing the vielbein basis $E^{\ul \mu}= e^{\lambda r}\,
dx^\mu$, $E^{\ul r} = dr$ (we are using a capital $E$ to denote the
vielbein, to avoid confusion with the exponential function). It follows that
the spin connection is given by 
\be
\omega^{\ul \mu \ul r} = \lambda\, E^{\ul \mu}\ ,\qquad \omega^{\ul \mu
\ul \nu}=0\ ,\label{spincon}
\ee
and the curvature 2-form is
\be
\Theta^{\sst{AB}}= -\lambda^2 E^{\sst A} \wedge E^{\sst B}\ .\label{curv}
\ee
Here, we are denoting tangent-space indices by $\sst A=(\ul \mu, \ul r)$,
{\it etc}.  Thus we see from (\ref{curv}) that the Riemann tensor has the
maximally-symmetric form $R_{\sst{ABCD}}=-\lambda^2 (\eta_{\sst{AC}}
\eta_{\sst{BD}} - \eta_{\sst{AD}}\eta_{\sst{BC}})$, as one expects since the
metric (\ref{ads}) describes \ads spacetime in $D$ dimensions, with Ricci
tensor $R_{\sst{AB}}=-\lambda^2 (D-1) \eta_{\sst{AB}}$. 

     The Killing spinor equations for a supergravity theory with cosmological
constant $\Lambda$ take the form
\be
\delta\psi_{\sst M} = D_{\sst M}\epsilon -
\ft12 \lambda \, \Gamma_{\sst M}\epsilon=0\ .\label{ks}
\ee
The constant $\lambda$ is the same one that we introduced previously, which
is related to $\Lambda$ by (\ref{lam}). Substituting the spin connection
(\ref{spincon}) for the \ads metric (\ref{ads}), we obtain the equations 
\be
\del_\mu \epsilon = \ft12 \lambda \, e^{\lambda r}\, 
\Gamma_\mu(1 + \Gamma_r)
\epsilon \ , \qquad
\del_r \epsilon =\ft12 \lambda\, \Gamma_r \,\epsilon\ ,
\ee
where $\Gamma_\mu$ and $\Gamma_r$ are understood to be the tangent-space
components of the $\Gamma$ matrices. We find that the solutions for the
Killing spinors $\epsilon$ are of two kinds, namely 
\bea
\epsilon &=& e^{\ft12\lambda r}\, \epsilon_+ \ ,\label{epp}\\
\epsilon &=& \Big( e^{-\ft12 \lambda r} +\lambda \, e^{\ft12\lambda r}
\, x^\mu \Gamma_\mu
\Big)\epsilon_- \ ,\label{epm}
\eea
where $\epsilon_\pm$ are arbitrary constant spinors satisfying
\be
\Gamma_r \epsilon_\pm =\pm \epsilon_\pm\ .\label{pm}
\ee
Thus in total, in $D$-dimensional \ads spacetime, we have $2^{[D/2]}$
independent Killing spinors.  Half of these, constructed using $\epsilon_+$,
are independent of the coordinates $x^\mu$, while the other half,
constructed using $\epsilon_-$, depend on $x^\mu$.  Note that this $x^\mu$
dependence is not periodic in nature.  It is interesting to note that all
the Killing spinors can be written in the single unified expression 
\be
\epsilon=e^{\ft12\lambda r \Gamma_r} \Big( 1+\ft12 \lambda\, x^\mu \Gamma_\mu
(1-\Gamma_r)\Big)\epsilon_0\ ,
\ee
where $\epsilon_0$ is an arbitrary constant spinor.  The Killing spinors
of 4-dimensional \ads spacetime were found in a different coordinate 
system in \cite{bf}.

     The explanation for the loss of half the Killing spinors upon
dimensional reduction of the $D$-dimensional \ads metric (\ref{ads}) to the
$(D-1)$-dimensional domain-wall metric is now clear:  The fields in $(D-1)$
dimensions that are retained in the Kaluza-Klein dimensional reduction are
those that are independent of the extra coordinate of the $D$-dimensional
spacetime.  In our case, we are taking this to be one of the spatial
world-volume coordinates $x^i$, since these are Killing directions in the
$D$-dimensional spacetime.  Indeed the Killing spinors (\ref{epp}) are also
independent of the coordinates $x^i$, and so these survive the reduction to
$(D-1)$ dimensions.  However, the other half of the Killing spinors, given
by (\ref{epm}), depend on the $x^i$ coordinates, and thus will not survive
the reduction.  In fact these Killing spinors depend linearly on the $x^i$ 
coordinates, and hence they would not survive even the periodic 
identification of the compactification coordinate, let alone the truncation 
to the zero-mode sector.

     This type of partial loss of supersymmetry at the level of solutions
can also be seen in the context of the usual $p$-branes.  Note that a
$p$-brane solution where the dilaton is regular on the horizon
(corresponding to cases where $a=0$, or its multi-charge generalisations)
interpolates between $D$-dimensional Minkowski spacetime at infinity, and
AdS$_{p+1} \times S^{\sst D -p-1}$ on the horizon.\footnote{It is
interesting to note that the ideal-gas entropy/temperature relation $S\sim
T^p$, which is satisfied by a non-dilatonic near-extremal $p$-brane
\cite{kt} or a regular-dilaton near-extremal $p$-brane \cite{lmpr}, is
related to the fact that the horizon of the $p$-brane is described by
AdS$_{p+1}\times S^{\sst D -p-1}$.} (This phenomenon has been observed in
the case of the $D=11$ membrane \cite{dgt} and in other cases \cite{bb}.) In
fact in these cases AdS$_{p+1}\times S^{\sst D -p-1}$ is also a solution of
the theory, which preserves all the supersymmetry (as, for example, in the
case of the AdS$_{4}\times S^7$ solution of $D=11$ supergravity).  However
Kaluza-Klein dimensional reduction of this solution, where the compactified
coordinate is one of the $x^i$ coordinates of the \ads metric (\ref{ads}),
gives rise to a domain-wall type solution in $(D-1)$-dimensions, which
preserves only half of the supersymmetry.  In fact, this can be viewed as a
higher-dimensional interpretation of the example we discussed previously,
where the supergravity theory with cosmological constant is now itself
viewed as coming from a yet higher dimensional supergravity compactified on
an appropriate sphere. 

\section{Elementary $(D-2)$-branes}

     The $(D-2)$-brane solutions that we have been discussing so far have
been solitonic, in the sense that they can be viewed as the degenerate
$n=0$ limit of $(D-n-2)$-branes using an $n$-form field strength with a
magnetic charge.  In general, one can consider an alternative formulation
of the theory in which the $n$-form field strength is dualised, to give
a $(D-n)$-form field strength.  The r\^oles of elementary and solitonic
$p$-branes are interchanged under this dualisation.  In the present context,
therefore, we may consider an alternative formulation of the theory with
a cosmological-type term, in which the Lagrangian (\ref{lag1}) is replaced
by
\be
{\cal L} = e R -\ft12 e (\del\phi)^2 -\fft1{2 D!}\, e\, e^{a\phi} F^2\ ,
\label{lag4}
\ee
where $F$ is a $D$-form field strength.  The equations of motion that follow
from this are essentially equivalent to those following from (\ref{lag1}),
except that now the constant $\Lambda$ has the interpretation of an
integration constant arising in the solution of the field equation for $F$. 
This has two important consequences.  Firstly, $\Lambda$ can now be zero,
allowing Minkowski spacetime as a solution of the theory.  Secondly, one can
allow $\Lambda$ to be only locally constant, taking different values in
different regions of the spacetime.  This permits more general classes of
domain-wall solutions, including configurations that describe multiple walls
in the spacetime \cite{bdgpt}.  Thus we may obtain solutions again given by
(\ref{dwmetric}), but where now $H$ can take the general form 
\be
H= c + b y + \sum_{i=1}^N m_i |y-y_i|\ .\label{multi}
\ee
These solutions describe $N$ domain walls located at the positions $y=y_i$. 
In the case $\Delta >0$, the curvature singularity that would occur at $H=0$
can be avoided by requiring, for example, $b=0$ and the constants $c$ and
$m_i$ to be all positive.  If instead $\Delta_{\rm AdS} < \Delta < 0$, the
curvature singularity at $H=\infty$, which was previously unavoidable in the
solitonic formulation in section 2, can now be avoided by requiring $b=0$
and that the constants $m_i$ in (\ref{multi}) satisfy $\sum_i m_i =0$. 
Finally, if $\Delta=\Delta_{\rm AdS}$ the solutions with $H$ given by
(\ref{multi}) describe portions of \ads spacetimes with different
cosmological constants that are joined together at domain walls, as 
described for $D=4$ in \cite{cdgs,gg,cgs}. 

     It should be remarked that the dualisation of the Lagrangian
(\ref{lag1}) to give (\ref{lag4}) is a subsector of the analogous process of
dualisation of an entire supergravity theory.  This has been carried out in
some detail in the case of the massive type IIA supergravity in $D=10$
\cite{bdgpt}. It is also worth remarking that in the dualised form, the
Kaluza-Klein reduction of the theory looks more conventional, since it now
admits a vacuum solution that is the direct product of $S^1$ and a Minkowski
spacetime \cite{bdgpt}.

\end{document}